\documentclass[%
reprint,
 amsmath,amssymb,
 aps,
]{revtex4-2} 
\usepackage{graphicx}
\usepackage{dcolumn} 
\usepackage{bm}

\begin{document}

\preprint{APS/123-QED}

\title{Multi-photon super-linear image scanning microscopy using upconversion nanoparticles}

\author{Yao Wang\textsuperscript{1,\#}}
\author{Baolei Liu\textsuperscript{1,\#}}
 \email{liubaolei@buaa.edu.cn}
\author{Lei Ding\textsuperscript{2}}
\author{Chaohao Chen\textsuperscript{3}}
\author{Xuchen Shan\textsuperscript{1}}
\author{Dajing Wang\textsuperscript{1}}
\author{Menghan Tian\textsuperscript{1}}
\author{Jiaqi Song\textsuperscript{1}}
\author{Ze Zheng\textsuperscript{1}}
\author{Xiaoxue Xu\textsuperscript{2}}
\author{Xiaolan Zhong\textsuperscript{1}}
\author{Fan Wang\textsuperscript{1}}
 \email{fanwang@buaa.edu.cn}
 
\affiliation{%
\textsuperscript{1}School of Physics, Beihang University, Beijing, 100191, China\\
\textsuperscript{2}School of Biomedical, Faculty of Engineering and IT, University of Technology Sydney, NSW 2007, Australia\\
\textsuperscript{3}Australian Research Council Centre of Excellence for Transformative Meta-Optical Systems, Department of Electronic Materials Engineering, Research School of Physics, The Australian National University, Canberra, ACT 2600, Australia\\
\textsuperscript{\#}These authors contributed equally to this work.
}%

\date{March 18, 2024}
             
\begin{abstract}
Super-resolution fluorescence microscopy is of great interest in life science studies for visualizing subcellular structures at the nanometer scale. Among various kinds of super-resolution approaches, image scanning microscopy (ISM) offers a doubled resolution enhancement in a simple and straightforward manner, based on the commonly used confocal microscopes. ISM is also suitable to be integrated with multi-photon microscopy techniques, such as two-photon excitation and second-harmonic generation imaging, for deep tissue imaging, but it remains the twofold limited resolution enhancement and requires expensive femtosecond lasers. Here, we present and experimentally demonstrate the super-linear ISM (SL-ISM) to push the resolution enhancement beyond the factor of two, with a single low-power, continuous-wave, and near-infrared laser, by harnessing the emission nonlinearity within the multiphoton excitation process of lanthanide-doped upconversion nanoparticles (UCNPs). Based on a modified confocal microscope, we achieve a resolution of about 120 nm, 1/8th of the excitation wavelength. Furthermore, we demonstrate a parallel detection strategy of SL-ISM with the multifocal structured excitation pattern, to speed up the acquisition frame rate. This method suggests a new perspective for super-resolution imaging or sensing, multi-photon imaging, and deep-tissue imaging with simple, low-cost, and straightforward implementations.
\end{abstract}

\maketitle

\section{Introduction} 
Fluorescence microscopy is an indispensable tool in biomedical sciences, but conventional techniques suffer from diffraction-limited resolution. The emergence of super-resolution microscopy circumvents this limitation and provides unprecedented nanoscale resolutions to visualize subcellular fine structures\cite{1}. The typical super-resolution methods include stimulated emission depletion (STED) microscopy\cite{2}, single-molecule localization microscopy\cite{3,4}, structured illumination microscopy (SIM)\cite{5,6,7}, and so on. These methods need either sophisticated excitation modulations or specific requirements for the used probes. Image scanning microscopy (ISM)\cite{8,9} offers a twofold enhancement in resolution with high signal-to-noise ratios, and is compatible with standard confocal laser scanning microscopes (CLSM) by simply replacing the single-point detector with a detector array. Recently, new modalities of ISM have been proposed for higher resolutions by combining with STED\cite{10}, photon antibunching\cite{11}, fluorescence fluctuations\cite{12}, or emission saturations\cite{13,14}. However, these variants need either modified optical configurations or additional requirements for the measurement. Thus, a simple and straightforward ISM for higher resolution is needed for practical applications.

Multiphoton fluorescence microscopy, on the other hand, provides deep penetration depth to observe the complex biological dynamics, by employing the near-infrared (NIR) transparent biological window. An implementation of ISM to the multi-photon imaging could double the resolution, such as the two-photon excitation ISM\cite{15,16} and the second-harmonic generation ISM\cite{17,18}. The multi-photon ISMs have limited resolution enhancement and need to use high-power, expensive femtosecond lasers. The recently emerged lanthanide-doped upconversion nanoparticles (UCNPs)\cite{19,20,21} attract enormous interest due to the low-power NIR excitation, tunable emission colors or lifetimes\cite{22,23,24}, excellent photostability, etc. UCNPs contain thousands of sensitizer ions (e.g., ytterbium and neodymium) and activator ions (e.g., erbium and thulium) to convert NIR excitation photons to shorter wavelength NIR, visible, and ultraviolet emission photons\cite{25}. They have been explored for super-resolution imaging, such as STED\cite{26,27,28}, deep-tissue imaging\cite{29}, nonlinear structured illumination microscopy (SIM)\cite{7}, multiplexed SIM\cite{30}, and point spread function (PSF) engineered imaging/classification\cite{31,32}. Most recently, the high-order optical nonlinearity (super-linear) exhibited in UCNPs offers new insight for developing the low-power, single-continuous-wave-beam super-resolution imaging with the simple confocal microscopes\cite{33,34}.

In this work, we present the multi-photon super-linear image scanning microscopy (SL-ISM) for the resolution enhancement with a factor exceeding two, by exploring the nonlinear fluorescence responses of UCNPs. By tuning the excitation NIR laser to a relatively low power, the upconverted emission with the highest nonlinearity leads to a much smaller emission PSF for the improved resolution in SL-ISM. Based on a modestly modified confocal microscope, we demonstrate SL-ISM with a resolution of about 120 nm, 1/8th of the excitation wavelength, by using ytterbium (Yb\textsuperscript{3+}) and thulium (Tm\textsuperscript{3+}) co-doped UCNPs. Furthermore, we expand SL-ISM with a multifocal illumination pattern for the faster frame rate across a large imaging area, facilitated by the integration of a microlens array. By introducing the super-linear fluorescence response into image scanning microscopy, our work provides a simple and low-cost multi-photon nonlinear ISM method by using only a single low-power, continuous-wave, near-infrared laser beam. 

\section{Results}

\begin{figure*}
\includegraphics[width=\textwidth]{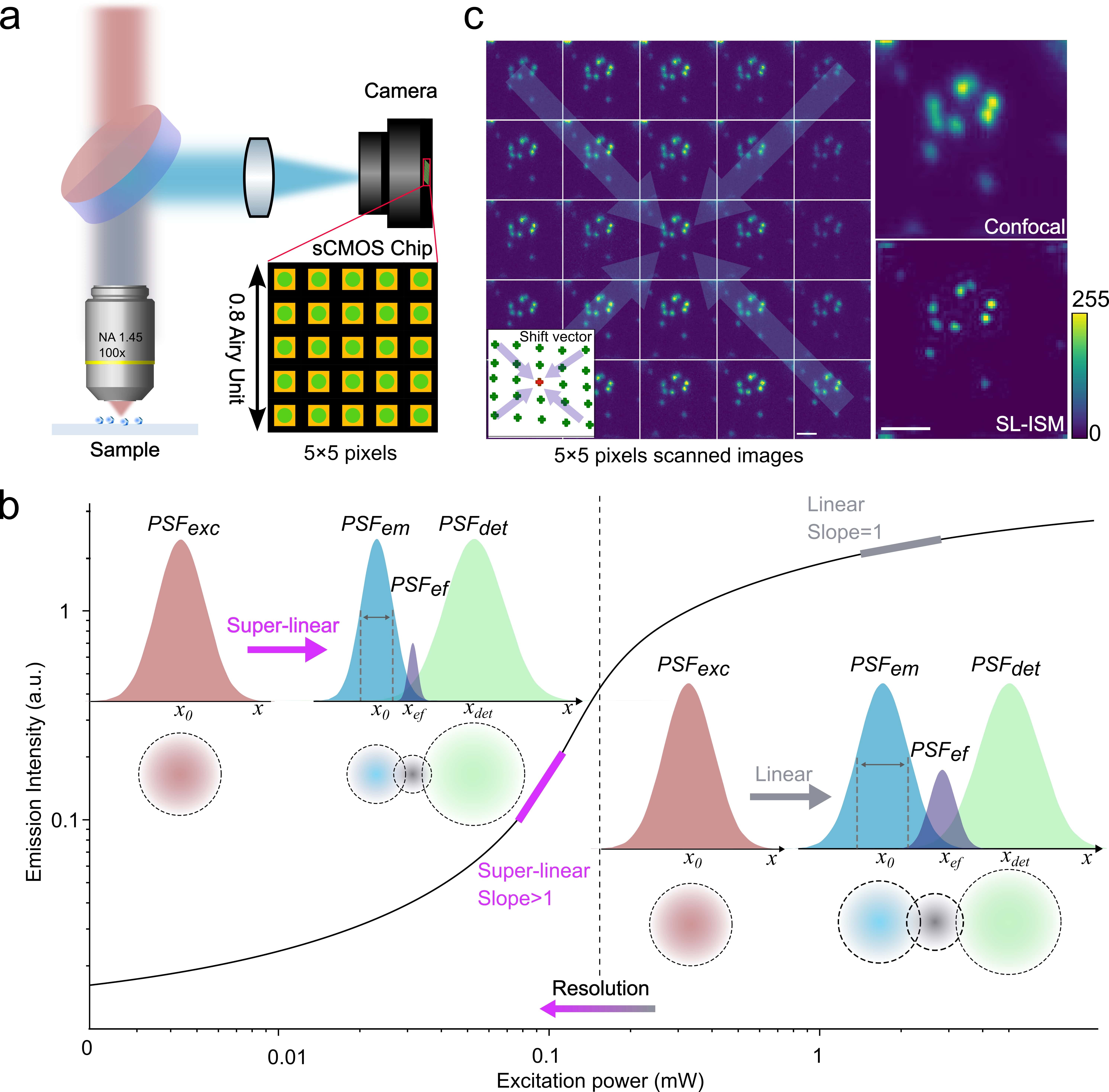}%
\caption{Principle and concept of multi-photon Super-Linear Image Scanning Microscopy (SL-ISM). (a) Optical setup of SL-ISM. The excitation beam (976 nm) passes through the dichroic mirror (DM) and is focused by the objective lens on the sample plane. The upconversion fluorescence (448±20 nm) from the sample is collected and imaged on the detector plane via a pinhole plane and a pair of relay lenses, which are not shown here. (b) Schematic of SL-ISM and ISM. The 455nm emission from the UCNPs shows a unique nonlinear curve with the increase of the excitation power. The linear slope, marked as the gray thick line on the right with an excitation power of about 2 mW, leads to a conventional linear ISM, in which the effective PSF (\textit{PSF\textsubscript{ef}}) is the product of excitation (or emission here) and detection PSFs (\textit{PSF\textsubscript{exc}}, \textit{PSF\textsubscript{em}}, \textit{PSF\textsubscript{det}}). With a lower excitation power of about 0.1 mW, the emission shows a strong super-linear dependence of the excitation power, which is marked as the purple thick line on the left. This super-linear slope would prompt a narrower \textit{PSF\textsubscript{em}}. Thus, the \textit{PSF\textsubscript{ef}} is further narrowed in SL-ISM with a much-enhanced resolution. (c) Schematic representation of SL-ISM reconstruction. The left shows the 5×5 matrix representing the images obtained from different detector pixels. The 25 images are firstly shifted to the center position by using adaptive pixel reassignment, with different shift vectors (bottom-left). Then, the Fourier reweight algorithm is used to enhance the resolution more. The comparison of SL-ISM and confocal images is shown on the right. The confocal image is obtained by summing the pixel intensities with an effective diameter of 0.8 Airy units (AU, at the emission wavelength). The scale bar is 500 nm.}
\label{fig:1}
\end{figure*}

Figure 1 presents the principle and concept of SL-ISM. The experimental setup is based on a home-built confocal microscopy setup but replacing the point detector with an array detector (Fig. 1a and Fig. S1), which is same as the traditional ISM. Here we use a near-infrared, continuous-wave 976 nm laser for the excitation and detect the upconverted 455 nm fluorescence. To illustrate the improved resolution induced by the super-linear fluorescence, we show the comparison of the typical linear ISM and SL-ISM in Fig. 1b. The unique nonlinear excitation-emission curve in Fig. 1b can be divided into super-linear range (with a slope of $>$ 1, marked as the purple thick line on the left), linear range (with a slope of 1, marked as the gray thick line on the right), and sub-linear range (with a slope of $<$ 1). For the linear range, the emission PSF ($PSF_{em}$) exhibits the same profile as the excitation PSF ($PSF_{exc}$). Thus, the effective PSF ($PSF_{ef}$) of the system in linear ISM that determines the spatial resolution is the product of $PSF_{em}$ (or $PSF_{exc}$) and the detection PSF ($PSF_{det}$), according to the theory of ISM\cite{35}, which is shown on the right of Fig. 1b. In the super-linear range, the fluorescence is emitted only at the center of the excitation beam ($PSF_{exc}$) leading to a narrower $PSF_{em}$ than the $PSF_{exc}$. Note that super-resolution imaging can be achieved by simply detecting the super-linear emission\cite{33}. For SL-ISM shown on the left of Fig. 1b, the $PSF_{ef}$ is the product of this super-linear emission induced narrower $PSF_{em}$ and the detection PSF ($PSF_{det}$), leading to a further improved resolution comparing to eiyher linear ISM or super-linear super-resolution imaging.

The reconstruction process of SL-ISM is similar to a traditional ISM, as shown in Fig. 1c. Each pixel of the array detector works as an inherent pinhole and an individual detector, to obtain a scanned image after laser scanning (Fig. 1c left). The images scanned by individual pixels with a position displacement of $d$ from the optical axis would have a $d/2$ position shift compared with the one aligned with the optical axis (Fig S3). That means the 5×5 pixel scanned images except the middle one are misaligned and they will have a position offset map, named shift vectors as shown in the bottom-left of Fig. 1c. So these pixels scanned images should be shifted to the central position according to the shift vectors, known as pixel reassignment\cite{9,36}. Then Fourier reweight algorithm\cite{8} is employed to further enhance the resolution and generate the SL-ISM image. The comparison of SL-ISM and confocal images is shown on the right of Fig. 1c, in which the former could resolve these isolated single nanoparticles while the latter could not.

\begin{figure*}
\includegraphics[width=\textwidth]{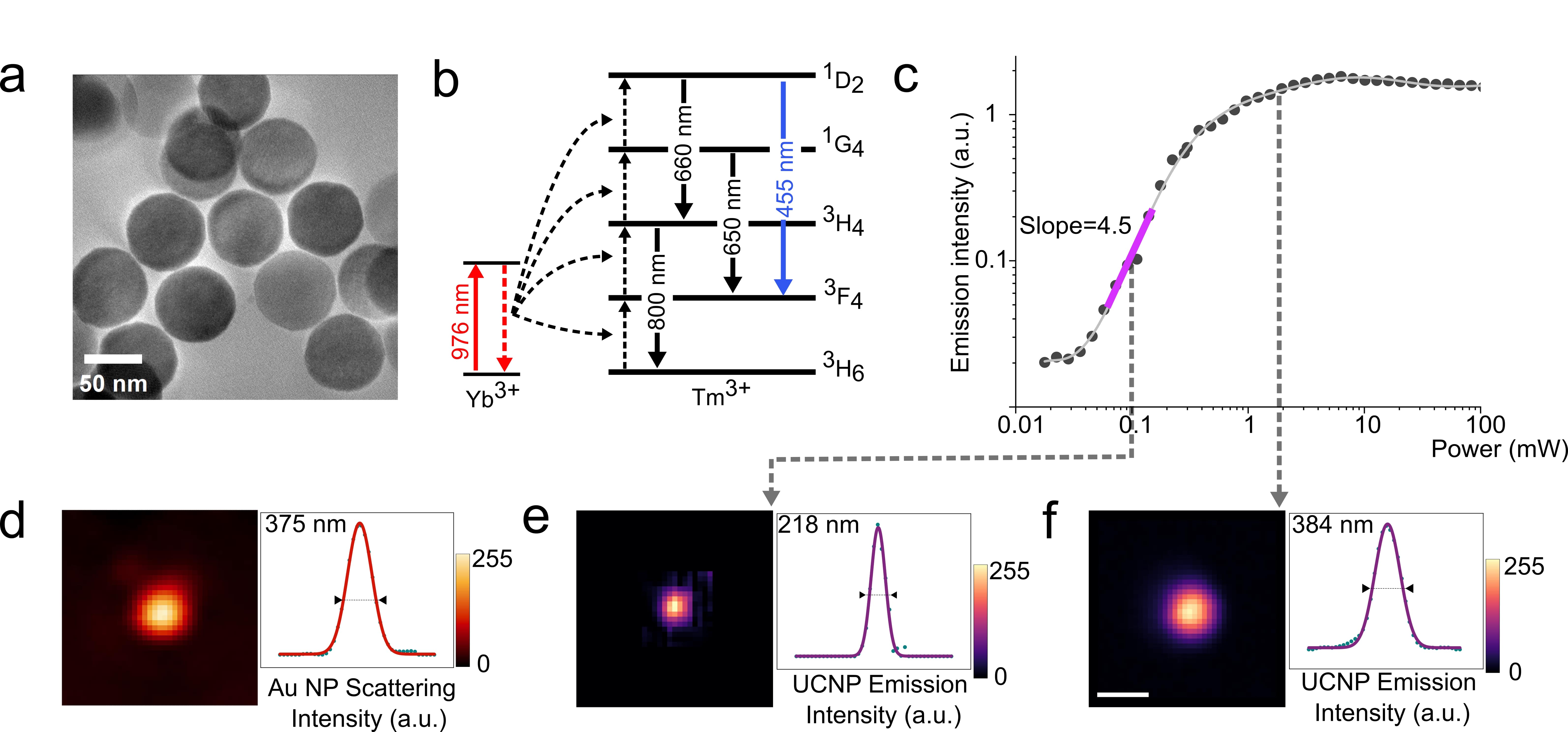}%
\caption{Super-linear nanoprobes and characterization of the resolution.(a) Transmission electron microscopy image of the used UCNPs (NaYF4: 20\% $\mathrm{Yb^{3+}}$, 8\% $\mathrm{Tm^{3+}}$), with an average size of around 68 nm. (b) The simplified energy level diagram for the upconversion processes in UCNPs. The 455 nm emission (five-photon process with the radiative transition from level 1D2 to 3F4 of Tm3+) is chosen for its strong nonlinearity. (c) Experimentally measured excitation-emission curve of a single UCNP, with a maximum slope of s=4.5 in the low-power range of about 0.1 mW. The slope decreases with the increase of the excitation power to s=1 at the power of 2 mW. (d) The diffraction-limit PSF of the excitation beam, with the FWHM of 375 nm ($\lambda/2.6$). The excitation PSF is measured by scanning a single isolated gold nanoparticle (50 nm) with a pinhole of about 0.55 AU. (e) The sub-diffraction imaging can be achieved by decreasing the excitation power in the super-linear range. The experimentally measured PSF has the FWHM of 218 nm ($\lambda/4.5$) with an excitation power of 0.1 mW. (f) In the high-power range, the measured PSF has an FWHM of 384 nm ($\lambda/2.5$), which is close to the diffraction limit in the linear region , with an excitation power of about 2 mW. The scale bar is 500 nm.}
\label{fig:2}
\end{figure*}

We then investigate the super-linear emission fluorescence in Fig. 2. Here the used nanoprobes are UCNPs of NaYF4: 20\% $\mathrm{Yb^{3+}}$, 8\% $\mathrm{Tm^{3+}}$ with a size of 68 nm (see Fig. 2a for the transmission electron microscopy image). Benefited from the multiple intermediate ladder-like energy levels, the NIR photon energy absorbed by $\mathrm{Yb^{3+}}$ sensitizer ions can be stepwise transferred onto different energy levels of $\mathrm{Tm^{3+}}$ activators (Fig. 2b). Multiple emission wavelengths of NIR, visible and ultraviolet are exhibited with different energy transition mechanisms, such as $\mathrm{{^3}H{_4}}$ → $\mathrm{{^3}H{_6}}$ (800 nm), $\mathrm{{^1}D{_2}}$ → $\mathrm{{^3}H{_4}}$ (660 nm), and $\mathrm{{^1}G{_4}}$ → $\mathrm{{^3}F{_4}}$ (650 nm). Here we choose the 455 nm emission (five-photon process of $\mathrm{{^1}D{_2}}$ → $\mathrm{{^3}F{_4}}$) due to its high nonlinearity and strong intensity. As shown in Fig. 2c, we measured the excitation-emission curve of a single UCNP at the 455 nm band by a custom-built confocal microscope with a 976 nm continuous-wave laser and a 100×, 1.45 NA objective lens. This unique excitation-emission curve showcases the steepest intensity evolution at the excitation power of about 0.1 mW, with a slope s = 4.5 (marked with a purple thick line). The slope then decreases as the increase of the excitation power, such as s = 1 at the excitation power of about 2 mW and s $<$ 1 at a higher power. 

To study the influence of the fluorescence nonlinearity on the system’s resolution, we performed the confocal imaging at two selected excitation powers, as shown in Fig. 2e\&f. To reveal the diffraction-limited resolution of the confocal system, we also measured the system’s excitation PSF by scanning the scattering light from an isolated gold nanoparticle (50 nm), which has the full width at half-maximum (FWHM) of 375 nm with an effective diameter of 0.55 Airy units (AU, at the scattering wavelength) (Fig. 2d), while the resolution of the confocal microscope with a fully-opened pinhole is expected to be about 410 nm that is close to the theoretical value\cite{33}. By taking advantage of super-linear fluorescence, the FWHM of the confocal image of a single UCNP at the excitation power of 0.1 mW decreases to 218 nm ($\lambda$/4.5) in Fig. 2e, which is almost twice better than the diffraction limit. By comparison, the measured FWHM increases to 384 nm in Fig. 2f, which is close to the diffraction limit, with about 2 mW excitation power of s = 1. 

\begin{figure*}
\includegraphics[width=\textwidth]{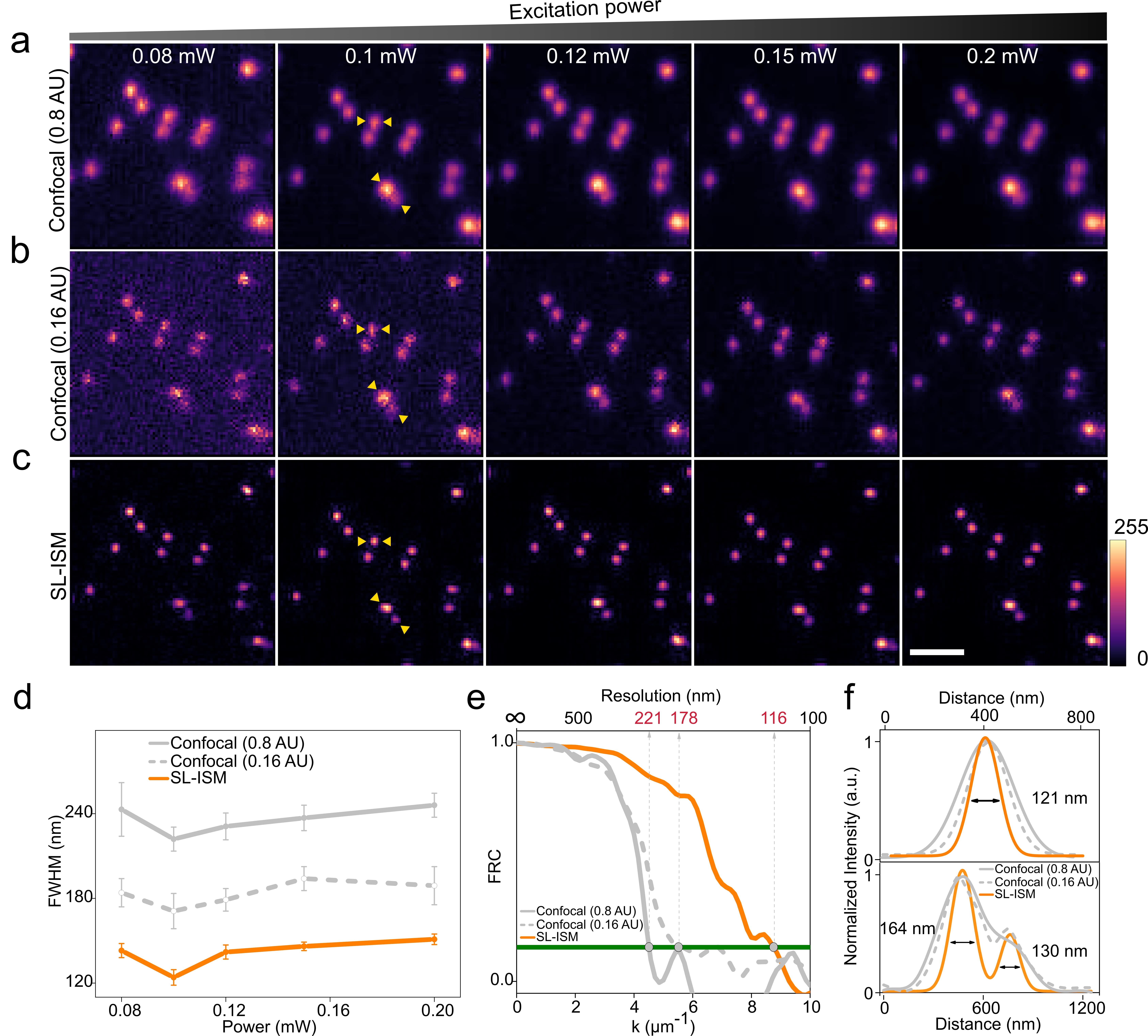}%
\caption{SL-ISM for super-resolution imaging. (a-c) Side-by-side comparison of confocal (0.8 AU), confocal (0.16 AU) and SL-ISM images of UCNPs at different excitation powers that range from 0.08 mW to 0.2 mW. The confocal (0.8 AU), confocal (0.16 AU) images are acquired by summing the intensities of the central 5×5 pixels or one pixel, respectively, of the camera’s pixel array. The SL-ISM images are reconstructed by using all the 5×5 pixels scanned raw images. SL-ISM images show both improved spatial resolutions and SNRs, compared with the other two methods. (d) The average FWHM of about 10 nanoparticles as the function of the excitation power. The resolutions for confocal (0.8 AU), confocal (0.16 AU) and SL-ISM are 222±8 nm, 171±12 nm, and 124±5 nm, respectively, corresponding to the excitation power of 0.1 mW with the strongest nonlinearity. (e) The FRC estimation curves of the three methods for the excitation power of 0.1 mW. (f) Normalized intensity cross-section profiles of images (a, b and c) along the indicated yellow arrows. The scale bar is 500 nm.}
\label{fig:3}
\end{figure*}

Based on the resolution enhanced by the super-linear fluorescence, SL-ISM further improves the spatial resolution as shown in Fig. 3. Here we demonstrate SL-ISM under variable excitation powers, as well as the comparison of confocal results, with the same UCNPs used in Fig. 2. The randomly distributed UCNPs are excited with the gradually varied laser power that ranges from 0.08 mW to 0.2 mW, which is near to the power range of the highest nonlinearity. The results obtained by three different modalities of confocal (0.8 AU), confocal (0.16 AU) and SL-ISM are presented in Fig. 3a, b, and c, respectively. The confocal (0.8 AU) images (Fig. 3a) have the worst spatial resolution. Due to the smaller pinhole, the confocal (0.16 AU) images (Fig. 3b) show better resolutions compared with the confocal (0.8 AU) results, but they suffer from more noise, especially when the excitation power is low. Compared to both two confocal methods, SL-ISM shows higher both resolution and image contrast/signal-to-noise ratio (SNR) to resolve the adjacent UCNPs, benefited from the pixel reassignment and Fourier reweight algorithms.

To quantify the resolution, we present the FWHM evolution curves as the function of the excitation power in Fig. 3d, based on statistics of about ten selected single nanoparticles. Both the three methods show the highest resolution at the power of 0.1 mW, which corresponds to the highest slope of the excitation-emission curve. Higher or lower excitation power deteriorates the resolution due to the slope’s decrease of the nonlinear excitation-emission curve. The averaged resolutions of confocal (0.8 AU), confocal (0.16 AU) and SL-ISM at 0.1 mW are 222±8 nm ($\lambda$/4\textsuperscript{th}), 171±12 nm ($\lambda$/6\textsuperscript{th}), and 124±5 nm ($\lambda$/8\textsuperscript{th}), respectively. The confocal (0.8 AU) or confocal (0.16 AU) methods have already achieved the sub-diffraction imaging, compared to the diffraction limit in Fig. 2. SL-ISM further improved the resolution by a factor of 1.8 or 1.4, compared to the other two methods. We also evaluate the resolutions of three methods at the 0.1 mW by Fourier ring correlation (FRC) analysis\cite{37}, which outputs resolutions of 221 nm , 178 nm and 116 nm for confocal (0.8 AU), confocal (0.16 AU) and SL-ISM, respectively. These results are consistent with FWHM-based resolutions (Fig. 3e). For better visualization, we show the Gaussian-fitted cross-section profiles of two selected areas in Fig. 3f, which are marked by yellow arrows in Fig. 3a-c. The line profile of a single nanoparticle from SL-ISM shows the best resolution, with the FWHM about 121 nm. Another example of two adjacent nanoparticles has shown SL-ISM can clearly resolve them, while the other two cannot. Furthermore, we derived the theoretical formula for the resolution enhancement with the increased nonlinearity slope in SL-ISM, which showcases the potential and superiority of SL-ISM compared with the super-linear super-resolution microscopy and linear ISM, as shown in Fig. S4. 

\begin{figure*}
\includegraphics[width=17.5cm]{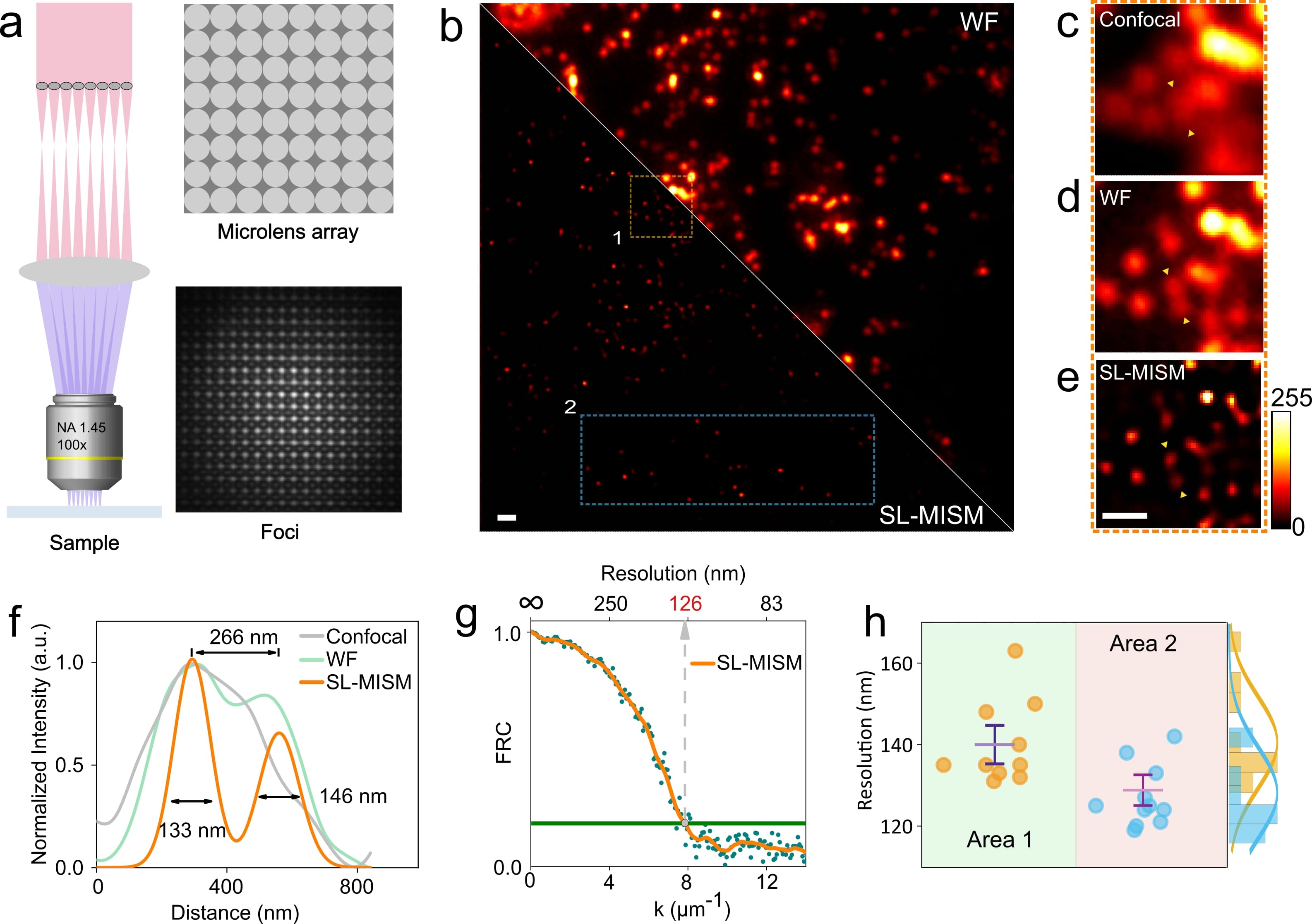}%
\caption{Experimental demonstration of multi-focal SL-ISM (SL-MISM). (a) Schematics of the SL-MISM setup. (b) Sub-diffraction SL-MISM (left) and wide-field (WF, right) images of 8\% $\mathrm{Tm^{3+}}$ doped UCNPs. (c-e) Comparison of (c) confocal, (d) WF, and (e) SL-MISM images for a selected area (the orange box) in (b). (f) Cross-section line profiles of two UCNPs marked by yellow arrows in (c), (d), and (e). (g) FRC estimation of the resolution for the SL-MISM image in (b). (h) Quantitative resolution analysis of the UCNPs located at two selected areas (the orange and cyan boxes) in (b). The Gaussian shape of the excitation beam induces the laser intensity differences of different areas in (b), which leads to a slight difference in the resolution due to different power-dependent excitation-emission nonlinearity. The averaged resolutions for the two areas are 139±10 nm and 128±8 nm, respectively, based on the measurement of about ten nanoparticles. The scale bar is 500 nm.}
\label{fig:4}
\end{figure*}

Furthermore, we extend SL-ISM with the multifocal illumination, named as SL-MISM, in a parallel detection scheme, to improve the frame rate of imaging acquisition for a large field of view (FOV). SL-MISM applies a microlens array to generate about 20×20 foci across a FOV of \(20~\mathrm{\text\textmu m}\)×\(20~\mathrm{\text\textmu m}\) (Fig. 4a), which would be 400-fold faster than SL-ISM with single-foci scanning. The SL-MISM’s result in Fig. 4b is acquired at 1 frame-per-second, with an exposure time of 4 ms for the raw data frame. A comparison of wide-field (WF) result is also present together, with a much lower resolution. Full images of them are available at the Supplementary Information Fig. S5. We show the comparison of confocal, WF, and SL-MISM in a small selected area (Area 1 in Fig. 4b, marked by the orange box) in Fig. 4c, d, and e. The Gaussian fitted profiles of two adjacent nanoparticles (marked with yellow arrows in Fig. 4c-e) in Fig.4f showcase the SL-MISM resolved them with a spacing of 266 nm and the FWHMs of 133 and 146 nm, respectively. As comparison, both confocal and WF methods cannot separate them (Fig. 4f). The resolution of the WF image is slightly higher than the confocal image, due to that the WF method directly detects the shorter emission wavelength while the confocal one scans the sample by the longer excitation wavelength. The FRC-based evaluation of the whole SL-MISM image shows a resolution of 126 nm (Fig. 4g, see further resolution analysis in Fig. S6).

We also show an interesting phenomenon of the slightly varied resolution in SL-MISM image in Fig. 4h. As described in Fig. 1, the resolution of SL-ISM is depended on the nonlinearity slope as well as the excitation power. The Gaussian-like excitation beam would generate the multi-foci with different excitation power densities (bottom-right of Fig. 4a) and thus lead to the resolution differences among the whole FOV. We analyze the differences by measuring the FWHMs from the center area (marked as the orange box in Fig. 4b) and the boundary area (marked as the cyan box in Fig. 4b) in Fig. 4h. The averaged FWHMs are about 139±10 nm and 128±8 nm, respectively. An optimization of the illumination pattern, such as the flat illumination field designed by the diffuser and microlens array-based system\cite{38}, could be conducted to achieve the highest and uniform resolution across the full FOV.

\section{Discussion}
Since the resolution of SL-ISM or SL-MISM is highly related to the nonlinearity of the excitation-emission curve, an optimization of the used probes with the higher nonlinearity slope would further improve the achieved resolution. The giant nonlinear response from photon-avalanche-based nanoparticles has a nonlinearity of 26th-order or higher can be used to realize a sub-70 nm resolution\cite{39,40}. An implement of these new nanoprobes in SL-ISM would help to push the resolution into sub-50 nm. The frame rate of SL-ISM can be increased by designing higher brightness UCNPs for shorter exposure time, using high-sensitive single-photon avalanche diode (SPAD) arrays\cite{36}, and adopting fast galvanometric mirrors or digital mirror devices (DMD)\cite{17}. The NIR emission from UCNPs could be explored to design the SL-ISM with both NIR excitation and emission wavelengths for deep-tissue super-resolution imaging\cite{7,29}. The continuous-wave, lower excitation power used in SL-ISM, compared with pulsed-lasers in traditional multi-photon microscopies, also makes it suitable for imaging living cells with a low-cost, low phototoxicity, and more stable modality. Emerging frontiers brought by UCNPs would also benefit SL-ISM for imaging with new functions, such as mapping the thermal field with a sub-diffraction resolution by involving the ratiometric nanothermometry\cite{41}.

In summary, we have demonstrated a multi-photon super-linear fluorescence-based ISM method (SL-ISM) that enhances resolution beyond linear ISM. The strong nonlinearity from the lanthanide-doped UCNPs is explored to realize a resolution of about 120 nm in a \(20~\mathrm{\text\textmu m}\)×\(20~\mathrm{\text\textmu m}\) FOV, with a frame rate of 1 Hz. A low-cost, continuous-wave laser is used for excitation at low power. We, therefore, anticipate that the work will open new opportunities for developing fast, cost-effective, multi-photon super-resolution microscopes.
\section{Materials and Methods}
\subsection{Experimental setup}
All experiments were conducted with a home-built imaging system, as shown in Fig. S1. A polarization-maintaining single-mode fiber-coupled 976 nm diode laser (BL976-PAG900, controller CLD1015, Thorlabs) is the laser source to excite UCNPs. Then the collimator lens expands the beam’s waist width to a diameter of about 8 mm to fully cover the front pupil of the objective lens. The microlens array with a pitch of \(150~\mathrm{\text\textmu m}\) and a focal length of 5.6 mm (MLA150-5C, Thorlabs) forms the multifocal scanning component, as shown in the dashed box of Fig. S1. The long-pass dichroic mirror (FF875-Di01-25×36, Semrock) with a cutoff wavelength of 875 nm is used to separate the excitation laser and emission fluorescence. The laser is focused by the objective lens (OL, UPlanSApo, 100×1.45 oil, Olympus) to excite UCNPs and the fluorescence is collected by the same objective. Then the fluorescence is captured by the sCMOS camera (400BSI V3, Dhyana) with a pixel size of \(6.5~\mathrm{\text\textmu m}\). The effective size of the camera size is about 60 nm corresponding to 0.16 AU (AU=(1.22$\lambda$)⁄NA, at the 455 nm emission wavelength) at the sample plane. The scanning step and pixel dwell time for SL-ISM are 50 nm and 50 ms, respectively, meanwhile 50 nm and 4 ms for SL-MISM (See more details in Supplementary Information).
\subsection{The image reconstruction processing}
The SL-ISM reconstruction has the same algorithm as the conventional ISM, which is based on pixel reassignment. The misalignment of the pixel and optical axis leads to the position shift of different scanned images recorded by pixels with different positions, as shown in Fig. 1 and Fig. S3. The Pixel reassignment algorithm aims to compensate for the position shift to keep the high resolution and SNR, while confocal images will be obtained by adding pixels scanned images directly without position compensation. Here, we employ the adaptive pixel reassignment algorithm which obtains the shift vectors without pre-calibrations or prior information by using the phase-correlation-based image registration algorithm which is based on the fast Fourier transform (FFT). The position shift vector $r_{(i,j)}$ between the test image ($g_{(i,j)}$, i,j are the coordinates in the 5×5 pixels detector plane) and reference image ($g_{(3,3)}$, the central pixel scanned image) can be estimated by:
\begin{equation}
r_{(i,j)}=FFT^{-1}\left(\frac{FFT(g_{(i,j)})FFT(g_{(3,3)})^*}{|FFT(g_{(i,j)})FFT(g_{(3,3)})^*|}\right)
\end{equation}
Then compensating the shift according to the shift vector for each pixel and then summing 5×5 images together will get the high-resolution image, while the SL-MISM image is reconstructed by the Fourier reweight and joint Richardson–Lucy deconvolution method (iteration=50) from the multi-frame dataset\cite{42}.
\bigskip
\begin{acknowledgments}
This work was supported by the Beijing Municipal Natural Science Foundation (1232027, 4212051); National Natural Science Foundation of China (U23A20481, 62275010, 62075004, 11804018); China Postdoctoral Science Foundation (2022M720347, 2022TQ0020); the International Postdoctoral Exchange Fellowship Program (YJ20220241, YJ20220037).
\end{acknowledgments}
\section*{Author contributions}
B. L., F. W. and Y. W. conceived the project. Y. W. and B. L. performed the measurements and analyses. Y. W., B. L. and F. W. prepared the manuscript. All the authors participated in the discussion and confirmed the final manuscript.  
\section*{Conflict of interest}
The authors declare no competing interests.
\section*{Supplementary information}
See Supplementary Information for supporting content.
\nocite{*}

\bibliography{ref}

\end{document}